\documentclass{jltp}
\usepackage{amssymb}
\renewcommand{\vec}[1]{{\bf #1}}

\runninghead{P. W\"olfle and A. Rosch}{Fermi liquid near a quantum critical point}
\input{tcilatex}

\begin{document}

\title{Fermi liquid near a quantum critical point\thanks{%
It is a pleasure to dedicate this paper to Hilbert von L\"ohneysen on the
occasion of his 60th birthday, and to wish him many more happy years of
activity in physics.}}
\author{P. W\"olfle$^1$ and A. Rosch$^2$}
\address{$^1$ Institut f\"ur Theorie der Kondensierten Materie, Universit\"at
Karlsruhe,\\
76128 Karlsruhe, Germany\\
$^2$Institut f\"ur Theoretische Physik, Universit\"at zu K\"oln, 50938
K\"oln, Germany\\
}
\maketitle

\begin{abstract}
We investigate the approach to the quantum critical point of a Pomeranchuk
instability from the symmetric, disordered side of the phase diagram. In the
low-temperature limit, a Fermi liquid description of the metal is possible
and becomes exact for $T \to 0$. We discuss in detail which features of the
approach to quantum criticality can be captured within Fermi liquid theory
and which are outside of its scope.

\end{abstract}


\section{Introduction}

Fermi liquid theory is the basis of our understanding of metals. In
recent years, quantum phase transitions in metals have been widely
studied mainly motivated by the experimentally observed 
(apparent) violation of Fermi liquid behavior for a wide variety of
systems\cite{hilbert,stewart,review}.  While a Fermi liquid
description may break down directly at a quantum critical point, it is
expected to be fully valid arbitrarily close to this singular point in
the low-temperature limit, $T \to 0$ (at least in the symmetric phase).
Therefore we want to explore in this paper the Fermi liquid
description of this regime.

From the point of view of Fermi liquid theory, the simplest and most
straightforward instability is one where the Fermi surface is deformed but
does not change its topology. Such Pomeranchuk\cite{pomeranchuk}
instabilities have been widely studied in the last few years \cite%
{oganesyan,metzner,wegner,chubukov,lawler,nilsson,schofield,kee}.
Experimentally, spontaneous deformations of correlated quantum systems, e.g.
of heavy nuclei or of crystalline lattices, have been well-known for
decades. More recently, purely electronic nematic transitions have been
suggested to be relevant for systems ranging from high-temperature
superconductors and 2-d Hubbard models \cite%
{oganesyan,metzner,wegner,kivelson}, Sr$_{3}$Ru$_{2}$O$_{7}$
close to its metamagnetic transition\cite{kee} to clean two-dimensional electron gases
in high Landau levels\cite{lilly}. Nevertheless, we are not aware of
experiments on quantum critical properties of a nematic transition e.g. from
a cubic to a tetragonal phase of a metal. An exciting option for the (near)
future is the controlled realization of a nematic transition of Fermions in
cold atom systems using Feshbach resonances in channels with higher angular
momentum. In these systems it should be possible to excite and measure
directly the relevant collective excitations by studying deformations of the
atom cloud in response to changes in the trapping potential.

Probably the most intriguing aspects of nematic phases in isotropic
systems is that deep in the ordered phase Fermi liquid theory is not
valid as pointed out by Oganesyan, Kivelson and
Fradkin\cite{oganesyan}. As explained above, we will, however, focus
our attention on the low-temperature behavior of the phase where the
symmetry is not broken. We first briefly outline the Fermi liquid
setup, then discuss the collective modes, the self-energy and finally
in the concluding section how Fermi liquid renormalizations couple
back to the order parameter dynamics.

\section{Fermi liquid model}

In the Fermi liquid phase of a metal fermionic
quasiparticles are the dominant low-energy excitations. At low temperatures
or, more generally, in weakly excited states the number of quasiparticles or
holes is small, providing a small expansion parameter. At the same time the
quasiparticle interaction may be strong, as measured in terms of
dimensionless Landau parameters $F_{\ell }^{J}$. For an isotropic system,
which we consider in the following, the single quasiparticle energy 
\begin{equation}
\epsilon _{k}=v_{F}(k-k_{F})  \label{1}
\end{equation}%
where $k_{F}$ is the Fermi momentum and $v_{F}=k_{F}/m^{\ast }$ the Fermi
velocity, is isotropic, and the Fermi liquid interaction $F_{k\sigma k{%
^{\prime }}\sigma {^{\prime }}}$ may be expanded\cite{fermiLiquid} in terms of spherical
harmonics, 
\begin{equation}
f_{k\sigma k{^{\prime }}\sigma {^{\prime }}}=\frac{1}{2N_F}\sum_{\ell
=o}^{\infty }\;\;P_{\ell }(\hat{k}\cdot \hat{k}{^{\prime }})[F_{\ell
}^{s}+F_{\ell }^{a}\sigma \sigma {^{\prime }}]  \label{2}
\end{equation}%
where $\sigma =\pm 1$ are the spin quantum numbers, $N_F$ the
quasiparticle density of states and $P_{\ell }(x)$ are in $d=3$
dimensions the Legendre polynomials, $P_{\ell }(\hat{k}\cdot
\hat{k}{^{\prime }})=\{4\pi /(2\ell +1)\}\Sigma _{m}Y_{\ell m}^{\ast
}(\hat{k})Y_{\ell m}(\hat{k}\prime )$, while in $d=2$ dimensions
$P_{\ell }(\hat{k}\cdot \hat{k}{^{\prime }})=\cos \ell (\phi -\phi
^{\prime })$ (with $\phi ,\phi ^{\prime }$ the angles of $%
\hat{k},\hat{k}{^{\prime }}$ with the x-axis). A change in the
quasiparticle occupation number $\delta n_{\vec{k}\sigma }$ leads to
the quasiparticle energy change
\begin{equation}
\delta \epsilon _{k\sigma }=\frac{1}{V} \sum_{k{^{\prime }}\sigma {^{\prime }}%
}f_{k\sigma k{^{\prime }}\sigma {^{\prime }}}\delta n_{k{^{\prime }}\sigma {%
^{\prime }}}  \label{3}
\end{equation}
where $V$ is the volume of the system.  A spatially non-uniform change
of occupation number with Fourier component $q$%
, $\delta n_{k\sigma }(q)$, will lead to a corresponding energy change
$%
\delta \epsilon _{k\sigma }(q)$. For sufficiently slow spatial
variation such that $q\ll k_{F}$, the Fermi wave number, the
dependence of $f_{k\sigma k{^{\prime }}\sigma {^{\prime }}}$ on $q$
may be neglected (it gives, however, a contribution\cite{metzner} to the
$q$ dependence of the order-parameter susceptiblity, discussed below).

On the microscopic level we will
adopt the model Hamiltonian\cite{fermiLiquid}
\begin{equation}
H=\sum_{k\sigma }\epsilon _{k}n_{k\sigma }+\frac{1}{2V}
\sum_{k,k{^{\prime }},q}' f_{k\sigma k{^{\prime }}\sigma {^{\prime }}}n_{k\sigma }(q)n_{k{%
^{\prime }}\sigma {^{\prime }}}(-q)\;\;,  \label{4}
\end{equation}%
where the prime on the summation sign indicates that the $q$-summation is
restricted to $q<q_{o}\ll k_{F}$.

In a crystal lattice the full rotation symmetry is reduced to the point
group symmetry. This implies that the Landau parameters $F_{\ell}^J$ do not
only depend on $\ell$, but also on $m$.

\section{Fluctuation propagator and Pomeranchuk instability}

The response of the system to an external field $\delta \epsilon _{k\sigma
}^{ext}(q,\omega )=\sum_{\ell m}Y_{\ell m}(\hat{k})$ $[\delta \epsilon
_{\ell m}^{ext,s}+\sigma \delta \epsilon _{\ell m}^{ext,a}]$, where $Y_{\ell
m}(\hat{k})$ are the spherical harmonics in $d=3$ (in $d=2$ dimensions the
expansion is in terms of \ $\sin \ell \phi ,\cos \ell \phi $), is expressed
for each channel $(\ell ,m,J=s,a)$ in terms of the generalized
susceptibility 
\begin{equation}
\chi _{\ell ,m}^{J}(q,\omega )=\frac{\delta n_{\ell m}^{J}}{\delta \epsilon
_{\ell m}^{ext,J}}\;\;.  \label{5}
\end{equation}%
We assume the vector $\vec{q}$ to be oriented along the z-axis in d=3 and
along the x-axis in d=2. Within RPA $\chi _{\ell ,m}^{J}$ is given by 
\begin{equation}
\chi _{\ell ,m}^{J}(q,\omega )=\frac{\Pi _{\ell ,m}}{1+F_{\ell }^{J}\Pi
_{\ell ,m}/2N_F(2\ell +1)}  \label{6}
\end{equation}%
where $\Pi _{\ell ,m}$ is the susceptibility in the absence of Fermi liquid
interactions, 
\begin{equation}
\Pi _{\ell ,m}(q,\omega )=-\sum_{\sigma }\int \frac{d^{3}k}{(2\pi )^{3}}4\pi
\;\Big|\;Y_{\ell m}(\hat{k})\Big|^{2}\frac{f(\epsilon _{k+q/2})-f(\epsilon
_{k-q/2})}{\omega +i0-\epsilon _{k+q/2}+\epsilon _{k-q/2}}  \label{7}
\end{equation}%
Expanding this expression in small $q$ and $(\omega /v_{F}q)$ one finds at
low temperatures    
\begin{equation}
\Pi _{\ell ,m}(q,\omega )=2N_F\Big[1-\xi _{0,\ell m}^{\prime 2}q^{2}+\beta
_{\ell m}^{\prime }(\omega,q)+i\gamma _{\ell m}^{\prime }(%
\omega,q)+O(q^{4})\Big]  \label{8}
\end{equation}%
where $N_F$ is the density of states (per spin) of quasiparticles  at the Fermi level. The
functions $\beta _{\ell m}^{\prime }$ and $\gamma _{\ell m}^{\prime
}$, discussed below,
depend on whether the considered mode is "even\textquotedblleft\ or
\textquotedblleft odd\textquotedblright, which means whether in $d=3$
dimensions $\ell +m$ is even or odd and in $d=2$ whether  the mode is 
 $\cos(2 n \phi)$ or $\sin(\{2 n+1\} \phi)$  or whether it is  
 $\sin(2 n \phi)$ or $\cos(\{2 n+1\} \phi)$.

 At this point one has to investigate which properties of the dynamic
 susceptibility $\chi_{\ell, m}^J(q,\omega )$ can be calculated within
 Fermi liquid theory. Both the $\omega =0$ response and the damping of
 the collective modes in the limit $\omega \rightarrow 0$ are
 low-energy properties which can be described exactly using Fermi
 liquid theory. However, this is not the case for the momentum
 dependence which gets contributions from high-energies, i.e. from the
 incoherent part of the excitation spectrum.  Especially, close to the
 critical point when the quasiparticle weight vanishes (see below),
 one can expect that the response is completely dominated by the
 incoherent part and a Fermi liquid calculation is impossible (further
 contributions also arise from momentum dependencies of the screened
 interactions). However, it is simple to give an order-of-magnitude
 estimate of this term by assuming that only renormalizations of order
 1 occur for this high-energy quantity. Taking into account that
 $N_F\propto 1/Z$ , where $%
 Z=m/m^{\ast }$ is the quasiparticle weight factor, and a relative
 factor of $%
 Z^{2}$ between the polarization $\Pi $ of quasiparticles (\ref{8})
 and electrons, we estimate
\begin{equation}
\xi _{0,\ell m}^{\prime 2}=\xi _{0,\ell m}^{2}\frac{1}{Z}\approx \frac{1}{%
k_{F}^{2}Z}\approx \frac{m^{\ast }}{k_{F}^{2}m}
\end{equation}%
fully consistent with results in the 
literature\cite{oganesyan,metzner,chubukov}.

In contrast, the damping term, $\gamma _{\ell m}^{\prime }$ , can be calculated rigorously from Fermi liquid theory,
yielding in $d=3$
\begin{eqnarray}
\gamma'_{\ell m}(\omega,q)=4\pi \Big|Y_{\ell m}(\cos \theta =%
\frac{\omega }{v_{F}q},\phi )\Big|^{2}\frac{\omega }{v_{F}q}
\end{eqnarray}
The coefficient of $\frac{\omega }{v_{F}q}$ in this expression is
finite in the limit $\omega /v_{F}q\rightarrow 0$, or of order
$(\omega /v_{F}q)^{2} $, depending on whether $\ell +m$ is even or
odd. This dependence arises because only quasiparticles with velocity perpendicular to
$\vec{q}$, i.e. with angle $\theta \approx \pi/2$, contribute to
Landau damping due to momentum and energy
conservation. This has the consequence that even and odd modes have
very different dynamical exponents. As for the even modes the Fermi
liquid renormalizations cancel between momentum and frequency
dependence, one can directly read off the dynamical critical exponent
$z_{%
  \mathrm{even}}=3$. To discuss $z_{\mathrm{odd}}$  we first have to
calculate the scaling of $m^{\ast }/m$ upon approaching the QCP which
is done in the next section.

The real part and the imaginary part of $\Pi _{\ell ,m}(q,\omega )$ are
connected by a Kramers-Kronig relation.\ In case of the even modes the
imaginary part does not give rise to a singular contribution to the real
part, i.e. $\beta_{\ell m}^{\prime even}$ may be neglected. However, for the odd modes, one
finds a contribution to the real part
\begin{eqnarray}
\beta _{\ell m}^{\prime odd}(\omega,q)=8\pi \Big|Y_{\ell
m}(\cos \theta =\frac{\omega }{v_{F}q},\phi )\Big|^{2}=b_{\ell m}\left(\frac{%
\omega }{v_{F}q}\right)^{2}
\end{eqnarray}

In $d=2$ dimensions the damping term is similarly suppressed whenever one of
the functions $\sin \ell \phi ,\cos \ell \phi $ vanishes at $\phi =\pi /2$.
Note that the suppression of damping is only effective for an isotropic
system. In a crystal lattice the eigenfunctions are pinned to the lattice
directions, and the condition of reduced damping can be only satisfied for
special directions of~$\vec{q}$.

In the limit of small $q$ and $\omega /v_{F}q$ we get finally 
\begin{equation}
\chi _{\ell ,m}^{J}(q,\omega )=\frac{2N_F(m/m^{\ast })}{\xi _{0,\ell
m}^{2}/\xi _{\ell m}^{2}+\xi _{0,\ell m}^{2}q^{2}-\beta _{\ell m}(
\omega,q)-i\gamma _{\ell m}(\omega,q)}  \label{10}
\end{equation}%
where the correlation length $\xi _{\ell m}$, the damping term $\gamma
_{\ell m}$ and the inertial term $\beta _{\ell m}$ are defined as 
\begin{eqnarray}
\frac{\xi _{0,\ell m}^{2}}{\xi _{\ell m}^{2}} &=&
\left(1+\frac{F_{\ell }^{J}}{2\ell +1}\right)%
\frac{m}{m^{\ast }}  \label{11} \\
\gamma _{\ell m}(\omega,q) &=&\frac{m}{m^{\ast }}\left\{ 
\begin{array}{ll}
\alpha _{\ell m}^{e}\, \omega /v_{F}q & ,  \ell +m\;\;\mathrm{even} \\ 
\alpha _{\ell m}^{o}(\omega /v_{F}q)^{3} & ,  \ell +m\;\;\mathrm{odd}%
\end{array}%
\right. \nonumber \\
\beta _{\ell m}(\omega,q) &=&\frac{m}{m^{\ast }}\left\{ 
\begin{array}{ll}
0  & ,  \ell +m\;\;\mathrm{even} \\ 
b_{\ell m}^{o}(\omega /v_{F}q)^{2} & ,  \ell +m\;\;\mathrm{odd}%
\end{array}%
\right. \nonumber
\end{eqnarray}%
in $d=3$ dimensions and
\begin{eqnarray}
\frac{\xi _{0,\ell m}^{2}}{\xi _{\ell m}^{2}} &=&\left(1+F_{\ell
  }^{J}\right)
\frac{m}{m^*} \label{d3xi} \\
\gamma _{\ell m}(\omega,q) &=& \frac{m}{m^*}\left\{ 
\begin{array}{ll}
\alpha _{\ell m}^{c}\, \omega /v_{F}q  & ,  \text{for }\cos \ell \phi \;,\ell
\;\mathrm{even}\text{; }\sin \ell \phi \;,\ell \;\mathrm{odd} \\ 
\alpha _{\ell m}^{s}(\omega /v_{F}q)^{3} & ,  \text{for }\sin \ell \phi
\;,\ell \;\mathrm{even;}\cos \ell \phi \;,\ell \;\mathrm{odd}%
\end{array}%
\right.  \nonumber \\
\beta _{\ell m}(\omega,q) &=& \frac{m}{m^*}\left\{ 
\begin{array}{ll}
0 & ,  \text{for }\cos \ell \phi \;,\ell \;%
\mathrm{even}\text{; }\sin \ell \phi \;,\ell \;\mathrm{odd} \\ 
b_{\ell m}^{s}(\omega /v_{F}q)^{2} & ,  \text{for }\sin \ell \phi \;,\ell \;%
\mathrm{even;}\cos \ell \phi \;,\ell \;\mathrm{odd}%
\end{array}%
\right. \nonumber
\end{eqnarray}%
in $d=2$ dimensions with numerical constants $\alpha_{lm}^{e,o,c,s}$. As we will show below, for even $\ell +m$ our
results for $\chi $ are fully consistent with the
literature\cite{oganesyan,metzner,chubukov}. Fermi liquid
renormalization effects are much more important for the odd modes (see
below),
which have either not been investigated\cite{metzner,chubukov} (they
are only relevant for uniform systems) or even when
considered\cite{oganesyan} the Fermi liquid corrections have not
been studied.

The uniform static susceptibility $\chi _{\ell m}^{J}(0,0)$ is seen to
diverge when 
\begin{eqnarray}
F_{\ell }^{J} &\rightarrow &-(2\ell +1),\text{ \ \ d=3}  \label{12} \\
F_{\ell }^{J} &\rightarrow &-1,\text{ \ \ \ \ \ \ \ \ \ \ \  d=2}
\end{eqnarray}%
signalling the Pomeranchuk instability\cite{pomeranchuk}. For $\ell =0$, $%
J=a $ this is nothing but the familiar ferromagnetic instability. We will,
however, only consider non-magnetic instabilities as ferromagnetic
transitions, where the Fermi surface splits, turn out to be substantially
more singular (Ref.~\onlinecite{chubukov} and references therein).


Near the quantum critical point the fluctuations induce an effective
dynamical quasiparticle interaction of the form 
\begin{eqnarray}
\Gamma _{kk{^{\prime }}}^{J}(q) &=&\sum_{\ell ,m}4\pi Y_{\ell m}(k)Y_{\ell
m}(k{^{\prime }})\Big(\frac{F_{\ell }^{J}}{2N_F}\Big)^{2}\chi _{\ell
m}^{J}(q,\omega )\;\;,  \label{13}
\end{eqnarray}%
whose effects are studied in the next section.

Note that for $d=3$ and even $\ell$ the transition is generically of first
order due to the presence of cubic terms in the Ginzburg Landau expansion
which are absent for odd $\ell$ in $d=3$ and for arbitrary $\ell$ in $d=2$.
Also a transition in the charge channel with $\ell=1$ is not possible for a
Galilean invariant systems as the ordered state with $\ell=1$ describes a
Galilei transformation which always costs energy\cite{schofield}. In the
following we will tacitly assume that a first order transition does not
take place (or is sufficiently weak).

\section{Self-energy}

The contribution to the quasiparticle self-energy induced by critical
fluctuations in channel $\ell $ takes the form 
\begin{eqnarray}
\func{Im}\Sigma (\vec{k},\omega ) &=&\Big(\frac{F_{\ell }^{J}}{2N_F}\Big)%
^{2}\int d\omega {^{\prime }}\;\int \frac{d^{3}q}{(2\pi )^{3}}\sum_{m}\Big|%
Y_{\ell m}({\bf \hat{k}}\cdot \hat{q})\Big|^{2}4\pi  \nonumber \\
&&\hspace{-6mm} \times \Big[n_{\omega {^{\prime }}}^{0}+f_{\omega {^{\prime }}+\omega }^{0}%
\Big]\;\;\func{Im}\{\chi _{\ell m}^{J}(q,\omega {^{\prime }-i0})\}\delta
(\omega {^{\prime }}+\omega -\epsilon _{\vec{k}+\vec{q}})  \label{14}
\end{eqnarray}%
where $n_{\omega }^{0}$ and $f_{\omega }^{0}$ are Bose and Fermi
distribution functions. We recall that the spherical harmonics are defined
with respect to the axis $\vec{q}$. Since $\func{Im}\chi _{\ell m}(q,\omega
) $ is isotropic in $\vec{q}$, the angular $\vec{q}$%
-integration removes the dependence on $\vec{k}$ , such that $\Sigma (\vec{k},\omega )$ is a function of 
$|\vec{k}|$ only, as expected for an isotropic system. At low temperatures, and small $\omega$ , $k$ 
the dominant contribution to the integral comes from small frequencies and by the energy conserving 
delta function, small ${\bf \hat{k}}\cdot \hat{q}$ . As discussed above, the 
quantity $4\pi \Big|Y_{\ell m}\Big|^{2}$ in this limit will be a constant $y_{\ell m}$ for even modes and equal to 
$x_{\ell m}({\bf \hat{k}}\cdot \hat{q})^{2}$ for odd modes.

The $\omega {^{\prime }}$- integration in Eq. (\ref{14}) may be readily performed. As
for the q-integration, we use the decomposition $\vec{q}=q_{r}{\bf \hat{k}}+\vec{q}%
_{t}$, where $\vec{q}_{t}$ is a $(d-1)$-dimensional vector in the plane
perpendicular to $\vec{k}$ and ${\bf \hat{k}}=\vec{k}/|\vec{k}|$. The
quasiparticle energy $\epsilon _{\vec{k}+\vec{q}}$ may be expanded in $q_{r}$
as 
\begin{equation}
\epsilon _{\vec{k}+\vec{q}}=(k+q_{r})v_{F}  \label{15}
\end{equation}%
where $v_{F}$ is the Fermi velocity. Then, 
\begin{eqnarray}
\func{Im}\Sigma (\vec{k},\omega )=&& \label{16}\\ 
\left(\frac{F_{\ell }^{J}}{2N_F}\right)^{2}&&\!\!\!\!\!\!\!\!\!\!\!\!
\int \frac{dq_{r}}{2\pi }\int \frac{d^{2}q_{t}}{(2\pi )^{2}}4\pi \Big|Y_{\ell m}\Big|^{2}\Big[%
n^{0}(\epsilon _{k+q}-\omega )+f^{0}(\epsilon _{k+q})\Big]\func{Im}\{\chi
_{\ell m}^{J}(q\;,\epsilon _{k+q}-\omega )\} \nonumber
\end{eqnarray}
At zero temperature the distribution functions reduce to step functions, e.g. for $\omega >0$ , 
\begin{equation}
n^{0}(\epsilon -\omega )+f^{0}(\epsilon )=\left\{ 
\begin{array}{lll}
-1 & , & 0<\epsilon <\omega \;\;\mathrm{or}\;\;-\omega <\epsilon <0 \\ 
0 & , & \mathrm{else}%
\end{array}%
\right.  \label{17}
\end{equation}%
and the contribution of even modes $\ell m$ to $\Sigma $ is then 
\begin{equation}
\func{Im}\Sigma (k,\omega )=\frac{(F_{\ell }^{J})^{2}}{2N_F}\int_{-k}^{%
\frac{w}{v_{F}}-k}\frac{dq_{r}}{2\pi }\;\int \frac{d^{2}q_{t}}{(2\pi )^{2}}%
\frac{y_{\ell m}\gamma (\widetilde{\omega },q)}{[\xi _{0}^{2}\xi ^{-2}+\xi
_{0}^{2}q^{2}-\beta (\widetilde{\omega },q)]^{2}+\gamma ^{2}}  \label{18}
\end{equation}%
where $\widetilde{\omega }=\omega -v_{F}(k+q_{r})$ . Here we dropped the
indices $\ell m$ of $\gamma _{\ell m},$ $\beta _{\ell m}$ and $\xi _{0,\ell
m}$ and introduced the correlation length $\xi $ of mode $\ell m$ by 
\begin{equation}
\xi ^{2}=\xi _{0,\ell m}^{2}(\frac{m^{\ast }}{m})\Big[1+F_{\ell }^{J}/(2\ell
+1)\Big]^{-1}\;\;.  \label{19}
\end{equation}%
It is seen that the main contribution to the $q$-integral comes from regions
where $q_{t}\gg q_{r}$, and therefore we may approximate $q\approx q_{t}$.

If $\ell +m$ is even, and $\gamma =\alpha \omega /v_{F}q$, $\beta =0$, one
finds with the new dimensionless variables $\tilde{q}_{r}=v_{F}(q_{r}+k)/%
\omega $ and $\tilde{q}_{t}=q_{t}/q_{\omega }$, where $q_{\omega }=(\alpha
\omega /v_{F}^{0}\xi _{0}^{2})^{1/3}$, and $v_{F}^{0}$ is the bare Fermi
velocity,

\begin{equation}
\func{Im}\Sigma (k,\omega )=\frac{(F_{\ell }^{J})^{2}}{2N_F}\frac{y}{%
\alpha }q_{\omega }^{d}\;\;\int_{0}^{1}\;\;\frac{d\tilde{q}_{r}}{2\pi }%
\;\int \;\frac{d^{d-1}q_{t}}{(2\pi )^{d-1}}\;\;\frac{\tilde{q}_{t}(1-\tilde{q%
}_{r})}{\tilde{q}_{t}^{2}(\zeta ^{2}+\tilde{q}_{t}^{2})^{2}+(1-\tilde{q}%
_{r})^{2}}  \label{20}
\end{equation}%
where $\zeta =(q_{\omega }\xi )^{-1}$. Here we have generalized the
expression to arbitrary dimension $d$.

Now the integration on $\tilde{q}_{r}$ may be performed to give 
\begin{equation}
\func{Im}\Sigma (k,\omega )=\frac{(F_{\ell }^{J})^{2}}{2N_F}\frac{y}{4\pi
\alpha }\;q_{\omega }^{d}\;\int \frac{d^{d-1}\tilde{q}_{t}}{(2\pi )^{d-1}}%
\tilde{q}_{t}\;\ell n(1+\frac{1}{Q^{2}})  \label{21}
\end{equation}%
where $Q=(\tilde{q}_{t}^{2}+\zeta ^{2})\tilde{q}_{t}$.

For small $\omega $, $\omega \ll \omega _{\xi }=v_{F}\xi _{0}^{2}/\xi
^{3}\alpha $, we have $\zeta \gg 1$ and $Q\approx \zeta ^{2}\tilde{q}_{t}$.
Using that $Q\gg 1$ in most of the integration regime, we finally get 
\begin{equation}
\func{Im}\Sigma (k,\omega )=\frac{y}{8\pi ^{2}\alpha }\;\;\frac{(F_{\ell
}^{J})^{2}}{2N_F}\Big(\frac{\alpha \omega }{v_{F}^{0}\xi _{0}^{2}}\Big)%
^{2}\;\xi ^{3}  \label{22}
\end{equation}%
Similarly one finds for a two-dimensional system 
\begin{equation}
\func{Im}\Sigma (k,\omega )=\frac{y}{8\pi ^{2}\alpha }\frac{(F_{\ell
}^{J})^{2}}{2N_F}\Big(\frac{\alpha \omega }{v_{F}^{0}\xi _{0}^{2}}\Big)%
^{2}\;\;\ell n\Big(\frac{v_{F}^{0}\xi _{0}^{2}}{\omega \alpha \xi ^{3}}\Big)%
\;\;\xi ^{4}  \label{23}
\end{equation}
consistent with results of Dell'Anna and Metzner\cite{metzner} if one
takes into account that in (\ref{23}) we have calculated the
self-energy of quasiparticles rather than the one of electrons which
is a factor $\frac{1}{Z}\approx \frac{m^*}{m}$ larger.

Within Fermi liquid theory one can calculate exactly the
quasi-particle decay rates in the limit of vanishing quasi-particle
energy as higher-order contributions beyond Eq.~(\ref{16}) are
supressed by phase-space factors. However,
it is generally {\em not} possible to
determine effective masses and the quasiparticle weight
$Z$ which are an input to rather than an output of Fermi liquid
theory. Close to the quantum critical point,
however, the singular corrections to real- and imaginary part of the self
energy come from the same critical fluctuations. Therefore we can
proceed by obtaining the real part of $\Sigma$ from Eq. (\ref{21}) by
replacing $\ell n(1+\frac{1}{Q^{2}})=\func{Im}\{+i\ell n(1+\frac{i}{Q})\}$
by $\func{Re}\{i\ell n(1+\frac{i}{Q})\}=-\mathrm{{arctan}\;\;\frac{1}{Q}}$.
It follows that 
\begin{equation}
\func{Re}\Sigma (k,\omega )=-\frac{y}{4\pi ^{2}}\frac{(F_{\ell }^{J})^{2}}{%
2N_F}\frac{\omega }{v_{F}^{0}\xi _{0}^{2}}\;\ln (\xi /\xi _{o}),\;\ \ \ \
\ \;d=3  \label{24}
\end{equation}%
which is still a quantity obtained from a quasiparticle description.
To calculate $Z\approx m/m^*$ we need the self-energy of the original
electrons $\Sigma ^{el}(k,\omega
)$ (rather than of the quasiparticles).  But this quantity can
be obtained using $\Sigma ^{el}(k,\omega
)=(m^{\ast }/m)\Sigma (k,\omega )$, to give
\begin{equation}
\frac{m^{\ast }}{m}=1-\frac{\partial Re\Sigma ^{el}}{\partial \omega }\Big|%
_{\omega =0}=1+\frac{1}{4}\Big(\frac{1}{k_{F}\xi _{0}}\Big)^{2}\Big(F_{\ell
}^{J})^{2}\ln (\xi /\xi _{o})  \label{25}
\end{equation}%
Only a logarithmic enhancement arises for  $\xi \rightarrow \infty $
in $d=3$.

In two dimensions, however, we find 
\begin{equation}
\func{Re}\Sigma (k,\omega )=-\frac{y}{32\pi }\;\frac{(F_{\ell }^{J})^{2}}{%
2N_F}  \frac{\omega}{v_{F}^{0}\xi _{0}^{2}}\;\xi  \label{26}
\end{equation}%
and hence 
\begin{equation}
\frac{m^{\ast }}{m}=1+\frac{1}{32}(k_{F}\xi _{0})^{-1}(F_{\ell
}^{J})^{2}(\xi /\xi _{o})  \label{27}
\end{equation}%
which is linearly diverging as $\xi \rightarrow \infty $.

In the case of odd $\ell +m$, where the additional contribution $\beta $
appears we may consider the limit $\gamma \rightarrow 0$, as $\gamma $ is a
subleading term compared to $\beta $, so that $\func{Im}\chi $ reduces to a
delta function. One finds  
\begin{eqnarray}
\func{Im}\Sigma (k,\omega )&=&\frac{(F_{\ell }^{J})^{2}}{2N_F}(\frac{m}{%
m^{\ast }})\;\int_{-k}^{\frac{w}{v_{F}}-k}\frac{dq_{r}}{2\pi }\;\int \frac{%
d^{d-1}q_{t}}{(2\pi )^{d-1}}x_{\ell m}(\frac{q_r}{q})^{2}\noindent \\
&&\pi \delta \lbrack \xi _{0}^{2}\xi ^{-2}+\xi
_{0}^{2}q^{2}-\frac{m^{\ast }}{m}(\frac{\widetilde{\omega }}{v_{F}^{0}q_{t}}%
)^{2}]  \label{28}
\end{eqnarray}%
At the Fermi surface ($k=0$), performing the $q_{r}$ - integration with the help of the delta function and
defining the dimensionless variable $p=q_{t}\xi $ one finds  
\begin{equation}
\func{Im}\Sigma (k,\omega )=\frac{\Omega _{d-1}}{4(2\pi )^{d-1}}\;\;\frac{%
(F_{\ell }^{J})^{2}}{2N_F}x_{\ell m}\sqrt{\frac{m^{\ast }}{m}}\;\;\frac{\xi
_{0}}{\xi ^{d+1}}\;\int_{0}^{\Omega _{c}}\;{dpp^{d-3}}(\Omega_{c}-p)^{2}\;\label{29}
\end{equation}%
where $\Omega _{c}^{2}=(m^{\ast }/m)(\omega /v_{F}^{0}\xi _{0})^{2}\xi ^{4}$
, and $\Omega _{d}$ is the surface area of the unit sphere in $d$ dimensions.
 We have used that in the limit of small 
frequencies we may neglect $p^{2}$ compared to $1$. In $d=3$ dimensions we find
\begin{equation}
\func{Im}\Sigma (k,\omega )=\frac{1}{24\pi }\;x_{\ell m}\;\frac{(F_{\ell }^{J})^{2}}{%
2N_F}\;\xi ^{2}\xi _{0}\;\Big(\frac{m^{\ast }}{m}\Big)^{2}\;\;\Big(\frac{%
|\omega |}{v_{F}^{0}\xi _{0}}\Big)\;^{3}\;\;\;\;.
\end{equation}%

In two dimensions one observes that the p integral is cutoff at the lower end 
by the requirement $q_t>q_r$ , yielding 
$p>p_c = (\frac{m^{\ast }}{m})(\frac{\omega \xi }{v_{F}^{0}})$. 
For odd $\ell +m$ one finds a contribution  
$\varpropto \omega ^{2}$, 
\begin{equation}
\func{Im}\Sigma (k,\omega )=
\frac{1}{8\pi }\;x_{\ell m}\;
\frac{(F_{\ell }^{J})^{2}}{2N_F}\;
\Big(\frac{m^{\ast }}{m}\Big)^{\frac{3}{2}}\;
\xi\;\;\;\Big(\frac{\omega}{v_{F}^{0}\xi_0}\Big)^2\;\; \ell n \;
\Big(\frac{m^{\ast}}{m}\frac{\omega}{v_{F}^{0}}\Big)
\;\;,  \label{30}
\end{equation}%
which is seen to be less divergent than the contribution from even modes.


In the conclusions we will discuss the effects of mass enhancement on
thermodynamics and the feed-back to the collective modes.

\section{Resistivity}

The contribution of critical fluctuations to the resistivity involves
the solution of a quantum kinetic equation. For a Galilean invariant
system the solution is simple: the center of mass of the charged
particles accelerates freely in the presence of an electric field
giving rise to a vanishing resistivity. Only if the momentum is
relaxed sufficiently rapidly due to impurities or Umklapp processes on
the Fermi surface in a lattice system, one can calculate the
resistivity from the transport relaxation rate.  There are two
contributions to this rate, from scattering-out processes and from
scattering-in processes. The former is characterized by a 
relaxation rate which is twice $\func{Im}\Sigma (\epsilon _{k},k)$.
The latter is an integral operator expressing the effect of vertex
corrections. It is well known that for an isotropic system the vertex
corrections reduce the probability for a collision in which momentum
$q$ is transferred by a factor $1-\cos \theta =(q/2k_{F})^{2}$.
Considering that the typical momentum transfer in a collision process
with a critical fluctuation is $\Delta q\sim \xi ^{-1}$ this reduction
factor is on average $\langle 1-\cos \theta \rangle \sim (k_{F}\xi
)^{-2}$.

Using these simple arguments, we expect for the transport scattering
rate of quasiparticles 
\begin{equation}
\frac{1}{\tau _{1}}\cong \Big(k_{F}\xi \Big)^{-2}\;\int \;d\epsilon _{k}\;2%
\func{Im}\Sigma (\epsilon _{f},k)\;\Big(-\frac{\partial f}{\partial \epsilon
_{k}}\Big)  \label{31}
\end{equation}%
and for the resistivity
\begin{equation}
\Delta \rho =\frac{m^{\ast }}{e^{2}n}\;\frac{1}{\tau _{1}}\;\;\;.  \label{32}
\end{equation}%
With $\int d\epsilon _{k}\func{Im}\Sigma (\epsilon _{k},k)\;\frac{\partial f
}{\partial \epsilon _{k}}\simeq -\func{Im}\Sigma (k_{F,}T)$, we obtain
from Eq.~(\ref{22}) in three dimensions (within the assumptions
specified above)
\begin{equation}
\Delta \rho \sim \frac{m}{e^{2}n}\;\frac{k_{F}^{2}}{m}\;\frac{(F_{\ell
}^{J})^{2}}{(k_{F}\xi _{0})^{3}}\;\Big(\frac{\xi }{\xi _{o}}\Big)\;\Big(%
\frac{T}{T_{F}^{0}}\Big)^{2}
\end{equation}%
where $T_{F}^{0}$ is the bare Fermi temperature, i.e. unrenormalized by
critical fluctuations (the noncritical renormalization is, however,
included). \ It is seen that $\Delta \rho $ increases linearly with the
correlation length $\xi ,$ and has the usual $T^{2}$ Fermi liquid
dependence.
In $d=2$ dimensions we obtain 
\begin{equation}
\Delta \rho \sim \frac{m}{e^{2}n}\;\frac{k_{F}^{2}}{m}\;\frac{(F_{\ell
}^{J})^{2}}{(k_{F}\xi _{0})^{2}}\;\Big(\frac{\xi }{\xi _{0}}\Big)^{2}\Big(%
\frac{T}{T_{F}^{0}}\Big)^{2}\;\ell n(T_{F}^{0}/T)  \label{34}
\end{equation}%
which depends quadratically on $\xi $ and has a logarithmic correction to
the $T^{2}$ temperature dependence.

\section{Conclusions}

Following Ginzburg and Landau, quantum phase transitions are most
often described in terms of order parameter fluctuations, i.e. bosonic
collective degrees of freedom. Recently, there has been an intense
debate whether such a description is in principle possible in metals
where always both bosonic and fermionic modes coexist at a quantum
critical point, for a recent review see Ref.~\onlinecite{review}. The
situation is much simpler in the regime considered in this article: in
the low-temperature limit of the symmetric phase, a Fermi liquid
description should always be possible, allowing for a controlled
calculation of low-energy properties. 
It is, however, not possible to calculate all physical quantities (like
the momentum dependence of the collective modes and the mass
renormalization) rigorously within the Fermi liquid approach as they
are influenced by high-energy modes. Nevertheless, with a few extra
assumptions on the regularity of the high-energy theory, we were able
to estimate these terms reliably.

Most of the results obtained in this paper are consistent with what is
expected from the purely bosonic theory. For example, the specific
heat coefficient $\gamma$ of a theory with dynamical critical exponent $z$ is
expected to be of the form\cite{hertz} $\gamma \sim T^{\frac{d-z}{z}}$
at the quantum critical point and accordingly $\gamma \sim \xi^{d-z}$
away from criticality for $T\to 0$ (with logarithmic corrections for
$d=z$). This is precisely what we obtained within the Fermi liquid
approach, see Eq.~(\ref{25}) and (\ref{27}).

Using the controlled Fermi liquid calculation one can check which
assumptions underlying a bosonic theory are valid. For example, for
the 'even' modes we find that all Fermi liquid corrections to the
order parameter susceptibility in (\ref{10}) and (\ref{11},\ref{d3xi})
cancel as $N_F/m^*$ and $m^* v_F$ are not renormalized. A similar
result was obtained by Chubukov, Pepin and Rech\cite{chubukov} who
found that there are no singular self-energy corrections to $\chi$ at
the $z=3$ critical point. Similarly, there are no singular Fermi
liquid corrections to the self-energy of quasiparticles
(\ref{22},\ref{23}) besides the trivial $Z$ factor, again consistent
with literature\cite{metzner,chubukov}. The situation is quite
different for the 'odd' modes where the $m^*/m$ corrections do not
cancel in Eqs.~(\ref{11}) and (\ref{d3xi}). This implies a breakdown of
the Hertz approach for these modes and changes, e.g., the lower
critical dimension of the Pomeranchuk quantum critical point which is
an interesting topic for future investigations\cite{unpublished}.

\section*{ACKNOWLEDGMENTS}
It is fitting to acknowledge that our interest in the problem of
quantum phase transitions has been kindled by the pioneering work of
Hilbert von L\"ohneysen. We are grateful to him for numerous
stimulating discussions and for delightful collaborations.
Furthermore, we thank M.  Garst, W. Metzner and M. Vojta for useful
discussions and acknowledge financial support by the DFG under SFB 608
(AR) and by the Helmholtz Virtual Institute for Quantum Phase
Transitions (PW).

\end{document}